\documentclass[aps, prd, 10pt, twocolumn, superscriptaddress,noshowpacs, preprintnumbers, longbibliography,nofootinbib,bibnotes,hyperref,floatfix]{revtex4-2}
\usepackage{tikz-feynman}
\usepackage{tikz}
\usepackage{amsmath}
\usepackage{amsfonts}
\usepackage{amssymb}
\usepackage{bbold}
\usepackage{epsfig}
\usepackage{graphicx}
\usepackage{bm}
\usepackage{array}
\usepackage{hyperref}
\usepackage{listings}
\usepackage{color}
\usepackage{float}
\usepackage{booktabs}
\usepackage{url} 
\usepackage[normalem]{ulem}

\usepackage{orcidlink}

\begin{document}

\title{Pauli blocking: probing  beyond-mean-field effects in  neutrino flavor evolution}

\author{Manuel Goimil-Garc\'ia \orcidlink{0009-0001-0518-9274}}
 \affiliation{Niels Bohr International Academy \& DARK, Niels Bohr Institute,\\University of Copenhagen, Blegdamsvej 17, 2100 Copenhagen, Denmark}
\author{Shashank Shalgar \orcidlink{0000-0002-2937-6525}}
\affiliation{Niels Bohr International Academy \& DARK, Niels Bohr Institute,\\University of Copenhagen, Blegdamsvej 17, 2100 Copenhagen, Denmark}
\author{Irene Tamborra \orcidlink{0000-0001-7449-104X}}
\affiliation{Niels Bohr International Academy \& DARK, Niels Bohr Institute,\\University of Copenhagen, Blegdamsvej 17, 2100 Copenhagen, Denmark}

\date{\today}

\begin{abstract}
Neutrino quantum kinetics in dense astrophysical environments is investigated relying on the mean-field approximation. In this paper, we heuristically explore whether beyond-mean-field effects due to neutrino degeneracy could hinder flavor instabilities that are otherwise foreseen. Our results show that these corrections shift the stability regions for a suite of (anti)neutrino ensembles: the flavor conversion of previously unstable distributions can be damped, but angular distributions that are stable in the mean-field case can also become unstable. Our work should serve as a motivation to further investigate the limitations of the mean-field treatment. 
\end{abstract}

\maketitle

\section{Introduction}

In a matter background, neutrino flavor evolution can be modified due to the refraction experienced by neutrinos~\cite{Wolfenstein:1977ue, Mikheyev:1985zog, Mikheev:1986if, Pantaleone:1992eq, Sigl:1993ctk}. A well-known example of this phenomenon is the Mikheyev-Smirnov-Wolfenstein effect: neutrino refraction on electrons can significantly alter the probability of flavor conversion~\cite{Wolfenstein:1977ue, Mikheyev:1985zog, Mikheev:1986if}. 
When the neutrino density is very large, such as in core-collapse supernovae and neutron-star merger remnants,   neutrino refraction due to other neutrinos leads to non-linear flavor evolution~\cite{Pantaleone:1992eq, Sigl:1993ctk, Mirizzi:2011tu,Tamborra:2020cul, Richers:2022zug,Volpe:2023met,Tamborra:2024fcd}.
Neutrino-neutrino refractive effects can make an incoherent mixture of neutrinos unstable~\cite{Banerjee:2011fj,Airen:2018nvp}, with small perturbations in the neutrino flavor states growing exponentially. A defining trait of neutrino self-interaction is that the evolution of all particles in the ensemble is coupled, leading to collective neutrino flavor conversion.

An interesting category of collective effects occurs when the (anti)neutrino momentum distribution is anisotropic. If the angular distributions of $\nu_{e}$ and $\bar{\nu}_{e}$ are such that the Electron Lepton Number (ELN) of neutrinos changes sign along a specific direction,  significant flavor evolution could take place, even in the limit of vanishing vacuum mixing term~\cite{Izaguirre:2016gsx,Chakraborty:2016lct,Morinaga:2021vmc,Padilla-Gay:2021haz,Fiorillo:2023mze,Fiorillo:2024dik,Dasgupta:2021gfs,Capozzi:2017gqd,Johns:2019izj}. 
In this case, the growth rate of flavor coherence is directly proportional to the neutrino density, for which reason these instabilities are named ``fast''~\cite{Sawyer:2005jk, Sawyer:2008zs, Sawyer:2015dsa}.

Fast flavor conversion is expected to occur in the neutrino decoupling regions of core-collapse supernovae and neutron-star merger remnants, with potential implications for the neutrino-driven explosion mechanism and the nucleosynthesis of  elements heavier than iron~\cite{Ehring:2023lcd,Ehring:2023abs,Nagakura:2023mhr,Wu:2017drk,George:2020veu,Just:2022flt,Fernandez:2022yyv,Li:2021vqj}. However, our current understanding of collective flavor conversion is based on solutions of neutrino quantum kinetic equations that employ the mean-field approximation~\cite{Sigl:1993ctk,Fiorillo:2024fnl}. The validity of the mean-field approach and the nature of the resulting phenomenology are the subject of ongoing investigation. For example, recent work attempts to look for many-body corrections to the mean-field treatment (cf., e.g., Ref.~\cite{Patwardhan:2022mxg} and references therein), but tractable systems are too idealized to give an insight on any eventual implications on astrophysical sources~\cite{Shalgar:2023ooi,Johns:2023ewj}.

\begin{figure}[t]
\centering
\includegraphics[width=8.7 cm]{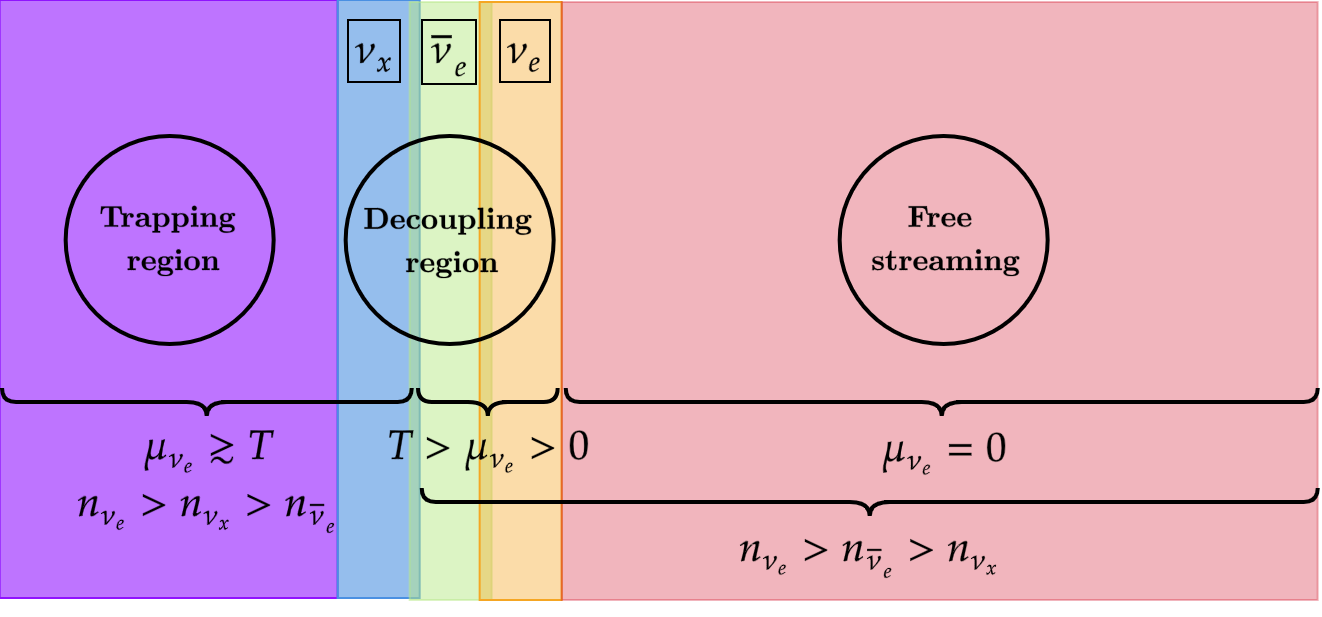}
\caption{Sketch of the neutrino decoupling region in the core of compact astrophysical sources. Close to the center, where the chemical potential of electron-flavor neutrinos is the highest, the local neutrino number density is isotropized by collisions with matter, and neutrino flavor conversion is not expected. 
Moving at larger distances from the source core, the neutrino degeneracy decreases, neutrinos gradually decouple from matter, and flavor instabilities are expected as ELN crossings develop.}
\label{fig:toon}
\end{figure}

The mean-field neutrino-neutrino Hamiltonian  describes  the interaction between a neutrino field, which represents  neutrinos with a certain momentum, and an external potential generated by other neutrinos. In  this picture, many-body correlations between particles with different momenta (``momentum correlations'') are explicitly forbidden. As a result, all neutrinos are in momentum eigenstates and scattering is only allowed in the form of a momentum swap. Momentum correlations would be most significant close to the core of compact astrophysical sources, where neutrinos are very degenerate (see the sketch in Fig. \ref{fig:toon}). We argue that said correlations could, in fact, depend on the shape of the neutrino spectrum, and that  degeneracy could therefore lead to novel effects, e.g.~flavor instabilities.

The region where beyond-mean-field corrections due to neutrino degeneracy could affect flavor evolution can be constrained \textit{a priori}. At nuclear densities, neutrinos of all flavors are in thermal equilibrium with matter because of frequent interactions with the background. The chemical potential for the heavy-lepton flavors ($\nu_x$ and $\bar{\nu}_x$) is negligible, and for electron (anti)neutrinos, $\nu_e$ ($\bar{\nu}_e$), it is positive (negative). This leads to the following hierarchy between the number densities of the different neutrino species: $n_{\nu_{e}} > n_{\nu_{x}} > n_{\bar{\nu}_{e}}$. Hence, as long as neutrinos are trapped, flavor 
instabilities are disfavored.
At larger distances from the source core, as the baryon density decreases and neutrino decoupling occurs,  $n_{\nu_{x}} < n_{\nu_{e}}, n_{\bar{\nu}_{e}}$; in this region, the (anti)neutrino angular distributions become forward peaked and ELN crossings develop~\cite{Tamborra:2017ubu, Brandt:2010xa, Shalgar:2019kzy}. This means that fast flavor instabilities can occur, but momentum correlations are unlikely to have an effect due to low neutrino density. If beyond-mean-field corrections caused by neutrino degeneracy should have an impact on flavor evolution, their effect should originate in the region of transition between the trapping and  free-streaming regimes. 

In this paper, we explore the impact of beyond-mean-field corrections due to neutrino degeneracy on neutrino quantum kinetics and, in particular, whether flavor instabilities can arise from or be suppressed by these effects. We stress that we do not aim to provide a self-consistent formalism to describe this many-body phenomenon. Our goal is  to  test whether many-body effects caused by  neutrino degeneracy could modify the development of flavor instabilities. Since the many-body equations of motion of neutrinos embedding this process are yet to be developed, we heuristically modify the single-particle equations of motion to assess whether dedicated formal work on the development of new neutrino equations of motion is justified. This work should serve as a testbed to eventually motivate further work to extend the commonly adopted mean-field approach. The corrections to the self-interaction Hamiltonian that we introduce are inspired by  the influence of the neutrino chemical potential on incoherent scattering. The neutrino emission rates are suppressed at energies $E\ll \mu_{\nu_l}$, because (some of) the final states in the neutrino phase space are already occupied: this effect is known as Pauli blocking. 

The manuscript is structured as follows. In Sec.~\ref{sec:mean-field}, we provide an overview of the neutrino equations of motion in the presence of neutrino self-interactions and introduce a prescription aiming to gauge beyond-mean-field effects linked to neutrino degeneracy. In Sec.~\ref{sec:lsa}, we linearize the neutrino equations of motion and, relying on a toy model,  compute the growth rate of flavor instabilities for a set of ELN angular distributions, within the single-energy approximation. 
In Sec.~\ref{sec:non-linear}, we present the solutions of the neutrino equations of motion for three benchmark ELN distributions. To understand the flavor evolution in the non-linear regime in the presence of neutrino degeneracy, we compare our results with the well-known case of the fast flavor pendulum~\cite{Padilla-Gay:2021haz,Johns:2019izj}. In Sec.~\ref{sec:multi-energy}, we relax the single-energy approximation and assess whether the energy dependence further affects the development of flavor instabilities. Finally, we discuss our findings and conclude in  Sec.~\ref{sec:end}. 
In addition, Appendix~\ref{sec:appendix} provides details on the correlations between neutrino creation and annihilation operators  in the presence of neutrino degeneracy.

\section{Neutrino equations of motion}
\label{sec:mean-field}

In the mean-field approximation, (anti)neutrinos are described by a set of $2\times 2$ density matrices of the form:
\begin{subequations}
\label{eq:rho}
\begin{alignat}{2}
\rho_{\mathbf{p}} &= 
\begin{pmatrix}
\rho_{ee;\mathbf{p}} & \rho_{ex;,\mathbf{p}} \\
\rho_{ex;\mathbf{p}}^* & \rho_{xx;\mathbf{p}}
\end{pmatrix} \ , \\
\bar{\rho}_{\mathbf{p}} &= 
\begin{pmatrix}
\bar{\rho}_{ee;\mathbf{p}} & \bar{\rho}_{ex;\mathbf{p}} \\
\bar{\rho}_{ex;\mathbf{p}}^* & \bar{\rho}_{xx;\mathbf{p}}
\end{pmatrix} \ ,
\end{alignat}
\end{subequations}
where $\rho_{ll';\mathbf{p}}$ and $\bar{\rho}_{ll';\mathbf{p}}$ denote the number density ($l=l'$) and flavor coherence ($l\neq l'$). These quantities are defined as correlation functions of two  creation and annihilation operators. Indeed, in a homogeneous medium, $\langle a^\dagger_{l',\mathbf{p'}}a_{l,\mathbf{p}}\rangle \equiv (2\pi)^3\delta (\mathbf{p}-\mathbf{p'})\rho_{ll';\mathbf{p}}$, where the angle brackets denote the average on the neutrino ensemble. If the properties of the ensemble are space dependent, then each matrix element can be defined locally by performing the Wigner transform of the correlation functions~\cite{Sigl:1993ctk, Stirner:2018ojk, Fiorillo:2024fnl}. The results then represent an average over a certain volume (momentum spread) around the position $\mathbf{r}$ (momentum $\mathbf{p}$). 
However, such an approach carries the assumption that inhomogeneities only occur at large scales, e.g.~due to temperature gradients in the medium.
Short-range fluctuations would be associated with large momentum uncertainties, which would make it impossible to specify $\mathbf{p}$. Moreover, in an inhomogeneous system, correlations between neutrino momenta would not be negligible, so the mean-field approximation would not hold in the first place~\cite{Volpe:2013uxl}. 

In this section, we first introduce the equations of motion of neutrinos within the mean-field approximation, and then modify them to take into account effects linked to neutrino degeneracy not included in the mean-field approach. We assume that the system is fully axisymmetric, so that the momentum distribution can be parametrized in terms of  energy, $E=|\mathbf{p}|c$, and  radial velocity $vc$. Moreover, we normalize the density matrices such that $ (2\pi )^{-1}\int \mathrm{d}^3p\mathrm{Tr}(\rho_\mathbf{p}) =1$ and $ (2\pi )^{-1}\int \mathrm{d}^3p\mathrm{Tr}(\bar{\rho}_\mathbf{p}) =\bar{n}_\nu /n_\nu$, where $n_\nu$ ($\bar{n}_\nu$) is the total (anti)neutrino number density. A factor $2\pi$ from the integral over the azimuthal angle is absorbed into the definition of $\rho_{\mathbf{p}}$ and $\bar{\rho}_\mathbf{p}$.

\subsection{Equations of motion within the mean-field approximation}

{Momentum correlations are not taken into account in the mean-field approximation. To first order in the Fermi constant, the evolution of the density matrices in Eq.~\eqref{eq:rho} is  determined by products of the form $\langle a^\dagger_a(\mathbf{p})a_{b}(\mathbf{q})\rangle \langle a^\dagger_{c}(\mathbf{p'})a_d(\mathbf{q'})\rangle$  (see Appendix \ref{sec:appendix} for details); this implies that the only contributions to the neutrino equations of motion are  momentum swaps between  (anti)neutrinos and annihilation and creation of neutrino-antineutrino pairs:}
\begin{subequations}
\label{eq:cfs}
\begin{alignat}{2}
{{\nu}}_a(\mathbf{p})+{\nu}_{b}(\mathbf{q}) &\to {\nu}_a(\mathbf{q})+{\nu}_{b}(\mathbf{p})\, ,\label{eq:cfs-1}\\
\nu_a(\mathbf{p})+\bar{\nu}_{a}(\mathbf{q}) & \to \nu_{b}(\mathbf{p}) +\bar{\nu}_{b}(\mathbf{q})\, ,\label{eq:cfs-2}
\end{alignat}
\end{subequations}
These processes are represented by the following Hamiltonian:
\begin{equation}
\label{eq:mean-field-H}
    H_\mathbf{p}=\mu\int \frac{\mathrm{d}^3p'}{2\pi E'^2} \left(1-\frac{\mathbf{p\cdot p'}}{|\mathbf{p}||\mathbf{p'}|}\right)(\rho_\mathbf{p'
    }-\bar{\rho}_\mathbf{p'})\, ,
\end{equation}
where  $\mu \equiv \sqrt{2}G_F (\hbar c)^3 n_\nu$ is the neutrino-neutrino interaction strength and $n_\nu$ is the total neutrino number density. We note that the mean-field Hamiltonian depends on the direction of the (anti)neutrino trajectory ($v$), but not on the energy of the particle. Accordingly, the mean-field equations of motion  for a homogeneous ensemble can be written as: 
\begin{subequations}
\label{eq:eom1}
\begin{alignat}{2}
i\hbar \frac{{\rho}_{\mathbf{p}}}{\partial t} &=\mu ([{D}_{0},{\rho}_{\mathbf{p}}] -v[{D}_{1},{\rho}_{\mathbf{p}}])\, ,\\
i\hbar \frac{\bar{\rho}_{\mathbf{p}}}{\partial t} &=\mu ([{D}_{0},\bar{\rho}_{\mathbf{p}}] -v[{D}_{1},{\bar{\rho}}_{\mathbf{p}}])\, ,
\end{alignat}
\end{subequations}
 where ${D}_n \equiv \int \mathrm{d}^3p'(2\pi E'^2)^{-1}L_n(v') ({\rho}_{\mathbf{p'}} -\bar{\rho}_{\mathbf{p'}})$. Here, $L_n$ are the Legendre polynomials, normalized such that $\int_{-1}^{+1} \mathrm{d}v' L_n(v')L_m(v') =(n+\frac{1}{2})^{-1}$, and $v'$ is the radial (anti)neutrino velocity.

 {Neutrinos can undergo different momentum-changing interactions. The interference among the different scattering amplitudes enhances the processes in Eqs.~\eqref{eq:cfs}. Equations~\eqref{eq:eom1}  represent the evolution of a homogeneous neutrino ensemble due to coherent scattering. If $a\neq b$, these amplitudes are indistinguishable from the flavor-changing amplitude due to vacuum neutrino mixing, so they can amplify the seed coherence provided by the latter--this is the mechanism responsible for collective flavor instabilities.}
 
{Recent literature has investigated whether  correlations may alter the mean-field description of flavor conversion; for example,  correlations could arise from incoherent scattering and damp collective oscillations due to quantum decoherence between the single-particle density matrices~\cite{Kost:2024esc}. 
Moreover, } neutrinos with the same flavor are indistinguishable, hence their quantum state has to be antisymmetric to particle exchange: {hence, momentum correlations should  inevitably appear  in a coherent many-body state.} When describing astrophysical environments, it is usually assumed that the system starts as a mixture of pure flavor eigenstates, i.e.~the density matrices are diagonal and the ensemble is fully incoherent. This assumption is well motivated: (anti)neutrinos in these systems are emitted as highly localized wavepackets with a specific flavor, which may not overlap. On the other hand, two plane waves with the same momentum always interfere. {The lack of momentum correlations  is thus not fully justified.} 

 \subsection{Beyond-mean-field effects linked to neutrino degeneracy}
\label{sec:bmf}
\begin{figure}[t]
\centering
\includegraphics[width=8.7cm]{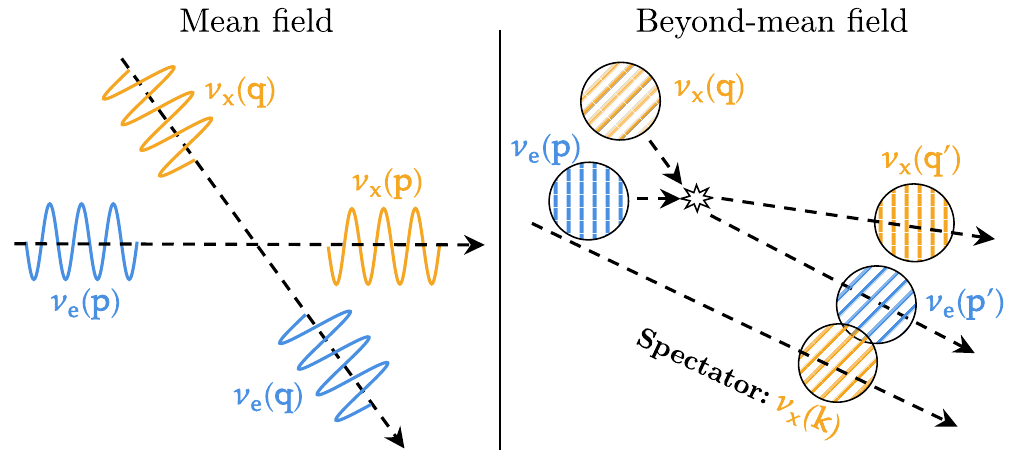}
\caption{{Illustrative} neutrino-neutrino interaction in the mean-field approximation (left) vs.~beyond-mean-field case (right). In both cases, {an electron neutrino (blue) scatters with a heavy-lepton-flavor neutrino (orange).} In the mean-field scenario, {neutrino plane waves swap their momenta: the process is indistinguishable from vacuum oscillations, and flavor coherence grows as a result.} In the beyond-mean-field picture, {there is a small recoil, so the final momenta are different, $\mathbf{q'\sim p}$ and $\mathbf{p'\sim q}$. As a result,} the incoming $\nu_e$ wavepacket is scattered into a region where there already are $\nu_x$'s with similar momentum ($\mathbf{k\sim p'}$), which do not participate in the scattering. {This interaction cannot affect flavor coherence, because $\nu_e(\mathbf{p'}\sim \mathbf{k})$ and $\nu_x(\mathbf{k})$ have to form a singlet state. We represent the neutrinos as flavor eigenstates for clarity, but in an astrophysical environment, they would be in flavor superposition due to vacuum mixing: hence, all of the particles that we sketch would be indistinguishable.}}
\label{fig:sketch-bmf}
\end{figure}

{For momentum correlations to vanish, the neutrino ensemble must be homogeneous over a certain length scale. Locally, neutrinos can then be described as plane waves, so their momentum is exact. The consequence, as shown in Eq.~\eqref{eq:cfs} and detailed in Appendix \ref{sec:appendix}, is that coherent processes only involve two momentum modes. In contrast, if momentum correlations are taken into account,} every neutrino contributing to the scattering amplitude can recoil to some extent. {The inhomogeneity implied by such correlations also means that} the momenta in Eq.~\eqref{eq:cfs} become approximate quantities. Hence, the interacting particles do not exactly swap their momenta, nor are they scattered precisely in the forward direction: rather, the initial and final momenta are indistinguishable within their uncertainty. This  momentum exchange could modify the development of flavor coherence (see Fig.~\ref{fig:sketch-bmf}): neutrinos cannot recoil into states that are already occupied, so certain coherent processes could be Pauli blocked if the neutrino ensemble is degenerate. {The suppression of coherent interactions would be described by momentum correlations between the interacting neutrinos and neighboring spectator neutrinos, representing the anti-symmetrization of the many-body neutrino quantum state.}

As a thought experiment, let us consider a collision between two beams of degenerate neutrinos, one made of $\nu_e$ and the other of $\nu_x$. At the mean-field level, the amplitude of $\nu_e\to \nu_x$ transitions would be the same as that of $\nu_x\to \nu_e$ processes. However, if momentum exchange is allowed, $\nu_{e}\to \nu_x$ should be favored in the first beam, and $\nu_x\to\nu_e$ in the other. In this case, the description of a coherent neutrino-neutrino interaction would require non-Hermitian operators, as it is done when incoherent scattering is taken into account~\cite{Sigl:1993ctk,1990Ap&SS.165...65R}. We note that the evolution of the system as a whole has to be Hermitian, but not for each momentum mode.

Working out the form of this non-Hermitian Hamiltonian exceeds the scope of this paper. Instead, we design a toy model encapsulating beyond-mean-field effects.

The occurrence of fast instabilities depends on the presence of a crossing in the ELN angular distribution~\cite{Morinaga:2021vmc,Fiorillo:2024bzm}. 
However, vacuum oscillations can relax this requisite: the mass difference term in the neutrino dispersion relation can be absorbed into the definition of an effective ELN angular distribution, with stability being determined by crossings in the latter~\cite{DedinNeto:2023ykt}. Likewise, we explore the possibility that beyond-mean-field effects in neutrino self-interactions can be described with effective density matrices, $\tilde{\rho}_\mathbf{p}$ and $\tilde{\bar{\rho}}_\mathbf{p}$. 

In the many-body approach, e.g.~if neutrino self-interactions were modeled through the  scattering between localized wavepackets, flavor evolution should be sensitive to the momentum uncertainty of each interacting wavepacket and  the spectator. Each neutrino would occupy a finite phase-space volume: thus, if an (anti)neutrino recoiled into the region occupied by an identical spectator neutrino, the overlapping wavepackets would be suppressed due to the exclusion principle. In a homogeneous single-particle description, quantum uncertainties are not defined;  the matrix elements $\tilde{\rho}_{ij;\mathbf{p}}$ and $\tilde{\bar{\rho}}_{ij;\mathbf{p}}$ can only depend on $\rho_{ll';\mathbf{q}}$ and $\bar{\rho}_{ll';\mathbf{q}}$. The following ansatz captures our main intuition that neutrino self-interaction cannot generate flavor coherence, if the (anti)neutrino phase space is fully occupied: 
\begin{subequations}
\label{eq:eom1-blocked}
    \begin{alignat}{2}
i\hbar \frac{{\rho}_{\mathbf{p}}}{\partial t}&=\mu ([\tilde{D}_{0},\tilde{\rho}_{\mathbf{p}}] -v[\tilde{D}_{1},{\tilde{\rho}}_{\mathbf{p}}])\, ,\\
i\hbar \frac{\bar{\rho}_{\mathbf{p}}}{\partial t}&=\mu ([\tilde{D}_{0},\tilde{\bar{\rho}}_{\mathbf{p}}] -v[\tilde{D}_{1},\tilde{\bar{\rho}}_{\mathbf{p}}])\, .
    \end{alignat}
\end{subequations}
Here
\begin{subequations}
\label{eq:rho_tilde}
\begin{alignat}{2}
    \tilde{{\rho}}_{\mathbf{p}} &= \left(\begin{array}{cc}
 [1-B_{e;\mathbf{p}}]^{1/2}\rho_{ee;\mathbf{p}} & \rho_{ex;\mathbf{p}}\\ 
 \rho_{ex;\mathbf{p}}^* & [1-B_{x;\mathbf{p}}]^{1/2}\rho_{xx;\mathbf{p}} \end{array}\right)\, ,\\
 \tilde{\bar{\rho}}_{\mathbf{p}} &= \left(\begin{array}{cc}
 [1-\bar{B}_{e;\mathbf{p}}]^{1/2}\bar{\rho}_{ee;\mathbf{p}} & \bar{\rho}_{ex;\mathbf{p}}\\ 
 \bar{\rho}_{ex;\mathbf{p}}^* & [1-\bar{B}_{x;\mathbf{p}}]^{1/2}\bar{\rho}_{xx;\mathbf{p}} \end{array}\right)\, ,
  \end{alignat}
 \end{subequations}
 and the effective Hamiltonian is written in terms of $\tilde{D}_n=\int \mathrm{d}^3p (2\pi E'^2)^{-1} L_n(v^\prime) (\tilde{\rho}_{\mathbf{p}^\prime}-\tilde{\bar{\rho}}_{{\mathbf{p}}^\prime})$. 
 
 The blocking functions $B_l$ ($\bar{B}_l$) represent the $\nu_l$ ($\bar{\nu}_l$) occupation numbers, 
 \begin{equation}
    B_{l;\mathbf{p}}  \equiv  \frac{(h c)^3}{2\pi E^2} n_\nu  \rho_{ll;\mathbf{p}}\ \mathrm{and}\  
    \bar{B}_{l;\mathbf{p}}  \equiv \frac{(h c)^3}{2\pi E^2} n_\nu  \bar{\rho}_{ll;\mathbf{p}}\, .
 \end{equation}
When the (anti)neutrino ensemble is in thermal equilibrium with matter, these functions are isotropic, following a Fermi-Dirac distribution: 
\begin{subequations}
\label{eq:blocking-fd}
\begin{alignat}{2}
    B^{\mathrm{eq}}_{l;E} &=\left[\mathrm{exp}\left({\frac{E-\mu_{\nu_l}}{T}}\right)+1\right]^{-1},\\
       \bar{ B}^{\mathrm{eq}}_{l;E} &=\left[\mathrm{exp}\left({\frac{E+\mu_{\nu_l}}{T}}\right)+1\right]^{-1},
    \end{alignat}
\end{subequations}
where $\mu_{\nu_l}$ is the chemical potential of neutrinos with flavor $l$, and $T$ is the local temperature. 

The form of the blocking factors $(1-B_{l;\mathbf{p}})^n$ is motivated by the effect of Pauli blocking on incoherent processes: scattering rates are suppressed by $\prod_i (1-B_i)$, where $i$ runs over all particles in the final state of the interaction. These factors come from the anticommutation relation of the neutrino creation and annihilation operators, $\lbrace a^\dagger_{l,\mathbf{p}}a_{l',\mathbf{p'}}\rbrace =(2\pi )^3\delta_{ll'}\delta (\mathbf{p}-\mathbf{p'})$~\cite{Sigl:1993ctk}. We choose the exponent  $n=1/2$ because it is not cross sections that enter the calculation of the refractive term, but scattering amplitudes.

In the rest of the paper, we investigate the effect of such beyond-mean-field corrections due to neutrino degeneracy. We stress that while the mean-field equations presented in Ref.~\cite{Sigl:1993ctk} do take into account Pauli blocking effects in neutrino-matter collisions, momentum correlations are not considered in the mean-field approach.

\section{Linear stability analysis}
\label{sec:lsa}

  \begin{figure*}
\centering
\includegraphics[width=18cm]{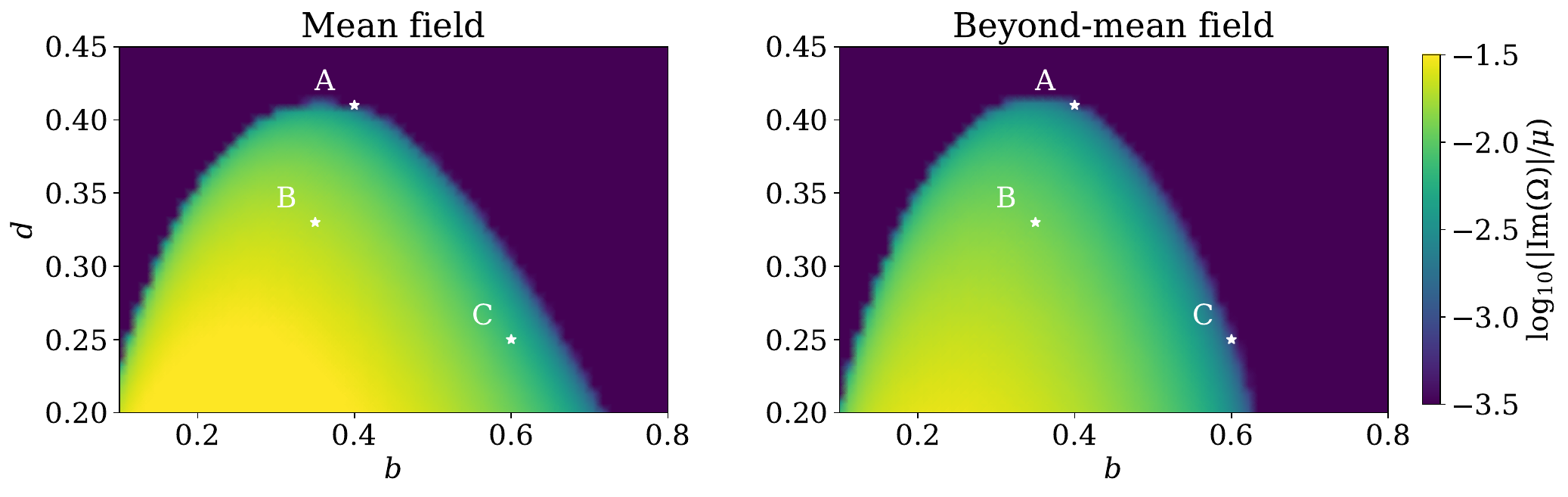}
\caption{Contour plot of the growth rate of the flavor instability in the plane spanned by the  parameters of the angular distributions $b$ and $d$, see Eq.~\eqref{eq:ang}, in the mean-field scenario (left panel) and for the case taking into account beyond-mean-field corrections due to Pauli blocking (right panel). Our three representative ELN distributions (Cases A, B, and C) are highlighted through the white stars. Overall, the beyond-mean-field prescription changes the stable regions of the ELN parameter space, with the growth rate being suppressed in most cases. 
}
\label{fig:hmap}
\end{figure*}

Given a set of initial conditions $D_\mathbf{p}(t=0)=(\rho_\mathbf{p}-\bar{\rho}_\mathbf{p})(t=0)$, the existence of flavor instabilities can be assessed by linearizing Eqs.~\eqref{eq:eom1} and solving the resulting eigenvalue problem, $i\dot{D}_{ex,\mathbf{p}}=\Omega D_{ex,\mathbf{p}}$~\cite{Airen:2018nvp}. In the single-energy approximation, where $E=|\mathbf{p}|c$ is fixed, the eigenfrequencies $\Omega$ are roots of the following polynomial: 
 \begin{equation}
        (I_0-1)(I_2+1) -I_1^2=0\, ,\label{eq:ch_poly}   
\end{equation}
with 
 \begin{equation}
        I_n=\mu\int \mathrm{d}v v^n\frac{\tilde{D}_{ee,\mathbf{p}}}{H_{ee,\mathbf{p}}-\Omega}\, ,
\end{equation}
where $H_{ee,\mathbf{p}}\equiv \tilde{D}_{0,ee}-v\tilde{D}_{1,ee}$ and  $D_{xx,\mathbf{p}}=0$. If $\Omega\in \mathbb{C}$, small perturbations in $D_{ex,\mathbf{p}}$ can grow exponentially at a rate $|\mathrm{Im}(\Omega)|$, and the ensemble  undergoes collective flavor conversion. 

We solve Eq.~\eqref{eq:ch_poly} for the following family of ELN distributions: 
\begin{subequations}
\label{eq:ang}
    \begin{alignat}{2}
    \rho_{ee;\mathbf{p}} &= \frac{1}{2}\frac{4\pi E^2}{(hc)^3}\frac{B^{eq}_{e;E}}{n_\nu}\, ,\\
    \bar{\rho}_{ee;\mathbf{p}} &= \rho_{ee;\mathbf{p}}\left[\frac{\bar{B}^{eq}_{e;E}}{B^{eq}_{e;E}} -\frac{F}{2} + \frac{0.2}{d}\exp \left( -\frac{(1-v)^2}{2b^2}\right)  \right]\, ,\\
    \rho_{xx;\mathbf{p}}& =\bar{\rho}_{xx;\mathbf{p}} =0\, ,
    \end{alignat}
\end{subequations}
where $F=0.2d^{-1}\int_{-1}^{+1} \mathrm{d}v \exp [-(1-v)^2/(2b)^2]$, and $d$ and $b$ are free parameters.
The occupation numbers $B^{eq}_{e;E}$ ($\bar{B}^{eq}_{e;E}$) explicitly depend on the (anti)neutrino energy, as well as on the thermodynamic properties of the medium. We consider $E =10.7$~MeV, $T=10$~MeV, and $\mu_{\nu_e}=3$~MeV, which may be representative of the $\bar{\nu}_e$ decoupling region in a core-collapse supernova at $\mathcal{O}(0.1)$~s post-bounce. Then, the functionals in  Eq.~\eqref{eq:ang} are such that electron neutrinos are isotropically distributed, whereas electron antineutrinos are forward peaked. We vary the parameters entering the angular distributions in the intervals $d\in [0.20,0.45]$ and $b\in [0.1,0.8]$, so that our parameter space encompasses both quasi-isotropic ELN distributions and functions with a very deep crossing.  Note that the neutrino energy dependence only enters the equations of motion through $B^{eq}_{e;E}$ and $\bar{B}^{eq}_{e;E}$, since we focus on fast flavor conversion and therefore the vacuum mixing term in the Hamiltonian is neglected, see  Eqs.~\eqref{eq:eom1-blocked}.

\begin{figure}[b]
\centering
\includegraphics[width=9cm]{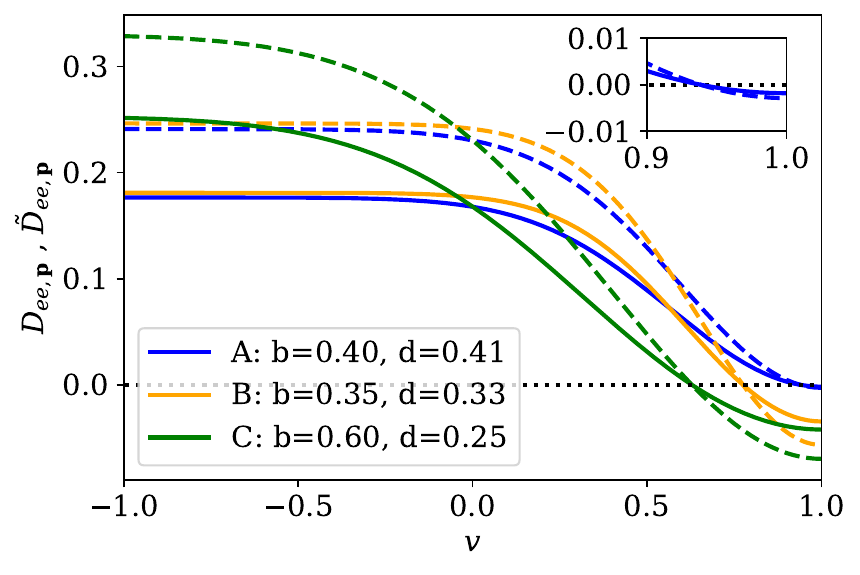}
\caption{Representative ELN angular distributions for Cases A, B, and C (cf.~Eq.~\ref{eq:ang} and Table~\ref{tab:growth}): $D_{\mathbf{p}}$ is represented by dashed lines and the effective $\tilde{D}_{\mathbf{p}}$ is plotted with solid lines. The inset highlights that there is an angular crossing in Case A, and its shape changes when beyond-mean-field corrections are taken into account, hence affecting the stability conditions. 
}
\label{fig:distr}
\end{figure}

Figure~\ref{fig:hmap} displays the growth rate of flavor instability, $\mathrm{Im}(\Omega)/\mu$, for our suite of ELN distributions in the plane spanned by the parameters $b$ and $d$ [see Eq.~\eqref{eq:ang}]. The mean-field (beyond-mean-field)  result is shown in the left (right) panel. In particular, we highlight three cases:  Case A, corresponding to $(b = 0.40, d = 0.41)$; Case B which has  $(b = 0.35, d = 0.33)$; and Case C obtained for $(b = 0.60, d = 0.25)$. These benchmark distributions are plotted in Fig.~\ref{fig:distr} (dashed lines). The growth rates for the flavor instabilities obtained in the mean-field and beyond-mean-field cases are summarized in Table~\ref{tab:growth} for the three selected distributions.

If the equations of motion are modified by beyond-mean field effects due to  Pauli blocking (cf.~Eq.~\eqref{eq:eom1-blocked}), the growth rate is suppressed. This is the case for Cases B and C, as shown in Table~\ref{tab:growth}. Nevertheless, for ELN distributions on the verge of becoming unstable, these same modifications can lead to a non-zero growth rate, see  Case A.  
These examples also illustrate that the growth rate of the flavor instability is not directly proportional to the depth of the ELN crossing, since $|\mathrm{Im}(\Omega_\mathrm{BMF})|$ is larger in Case A than in Case C. Rather, the stable region of our parameter space is determined by the behavior of the integrals $I_n$ defined in Eq.~\eqref{eq:ch_poly}, which depend not only on the shape of the crossed region, but also on the total lepton number ($D_0$) and flux ($D_1$)~\cite{Fiorillo:2023hlk,Padilla-Gay:2021haz}.

\begin{table}[H]
\caption{Growth rate of the flavor instability for our three selected angular distributions in Fig.~\ref{fig:distr}, in the mean-field   [$\mathrm{Im}(\Omega_{\rm{MF}})$] and beyond-mean-field [$\mathrm{Im}(\Omega_{\rm{BMF}})$] cases. 
}
\centering
\begin{tabular}{c|c|c|c}
\multicolumn{1}{c}{Distribution} &\multicolumn{1}{c}{$(b,d)$} & \multicolumn{1}{c}{$\mathrm{Im}(\Omega_\mathrm{MF}$) [$ \mu$]} & \multicolumn{1}{c}{$\mathrm{Im}(\Omega_\mathrm{BMF})$ [$\mu$]}\\ \toprule
Case A & $(0.40,0.41)$ & 0 & 0.0011 \\
Case B & $(0.35,0.33)$ & 0.0161 & 0.0108\\
Case C & $(0.60,0.25)$ & 0.0067 & 0.0008 \\ \bottomrule
\end{tabular}
\label{tab:growth}
\end{table}

\section{Non-linear flavor evolution}
\label{sec:non-linear}
To assess the effect of the modifications induced by neutrino degeneracy on flavor conversion,  we solve Eqs.~\eqref{eq:eom1-blocked}  for the angular distributions in Fig.~\ref{fig:distr}. Instead of generating flavor coherence over time through  vacuum mixing, we  consider  a small off-diagonal perturbation in the off-diagonal terms of the density matrices 
($\rho_{ex,\mathbf{p}}= 10^{-6}$ and $\bar{\rho}_{ex,\mathbf{p}}= 10^{-6}$). 

 \begin{figure*}
\centering
\includegraphics[width=\textwidth]{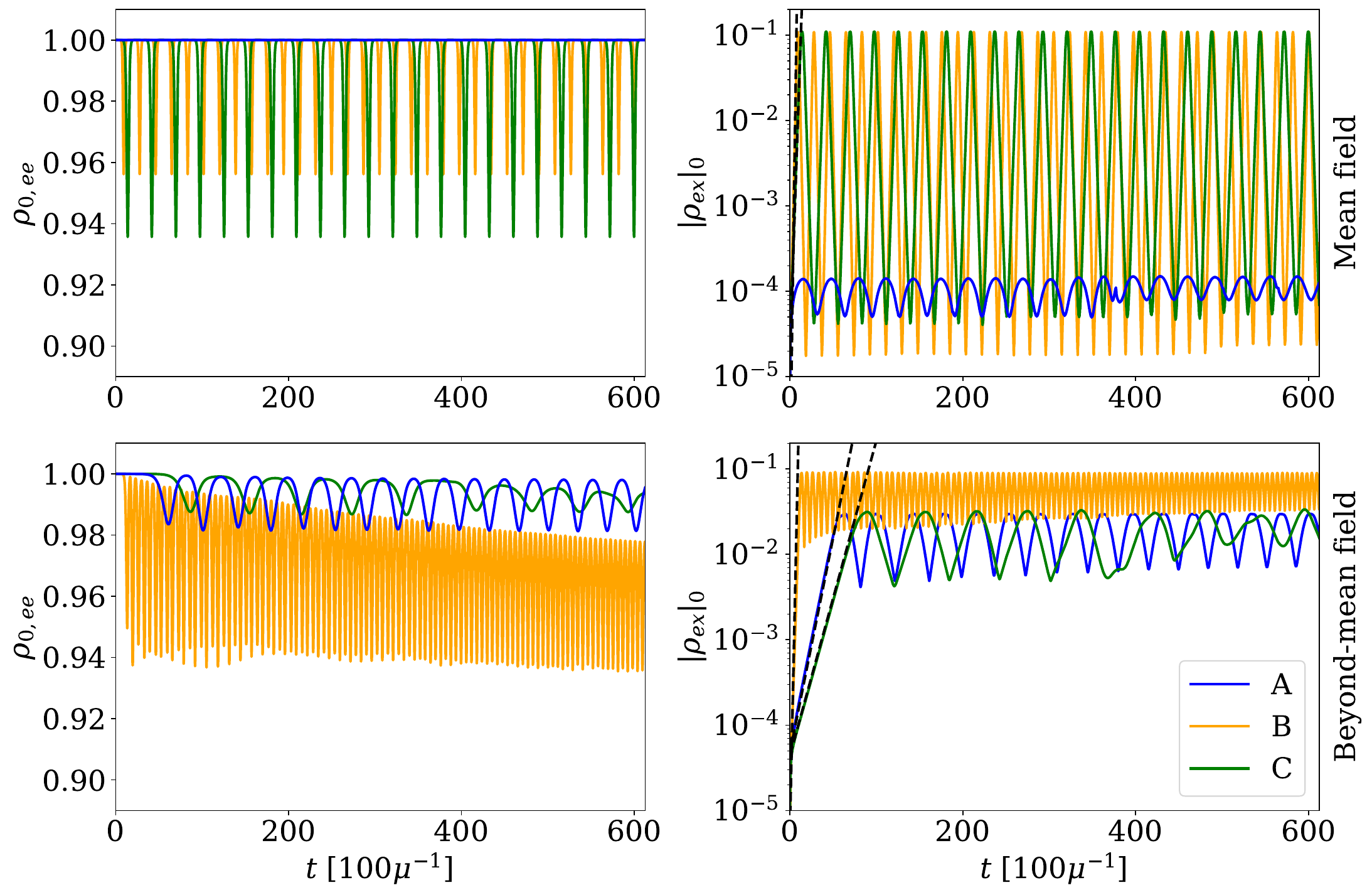}
\caption{Angle-integrated neutrino density (on the left) and flavor coherence (on the right)  as  functions of time for Cases A, B, and C (cf.~Fig.~\ref{fig:distr} and Table~\ref{tab:growth}). The top panels have been obtained in the mean-field scenario, while the bottom panels represent the flavor outcome when beyond-mean-field effects due to neutrino degeneracy are taken into account.
The results of the linear stability analysis are shown as dashed black lines.
The beyond-mean-field terms are responsible for making Case A unstable, and damping flavor conversion for Cases B and C.}
\label{fig:R-ee}
\end{figure*}
Figure~\ref{fig:R-ee} shows the angle-integrated neutrino density (left panels) and flavor coherence (right panels), $\rho_{0,ee}$  and $|\rho_{ex}|_0$, for Cases A, B, and C in the mean-field (top panels) and beyond-mean-field (bottom panels) scenarios. The linear growth rate of $|\rho_{ex}|_0$ agrees with the results from the normal mode analysis presented in Sec.~\ref{sec:lsa}.  Case A becomes unstable once the beyond-mean-field modifications are taken into account (note that the small oscillations visible on the top right panel are due to the initial seed).  For Cases B and C, an overall smaller amount of flavor conversion takes place when beyond-mean-field corrections are taken into account.

In order to better grasp our findings in the non-linear regime, it is useful to express the density matrices in the orthonormal basis $\lbrace 1_{2\times 2}, \sigma_x,\sigma_y ,\sigma_z\rbrace$, where $1_{2\times 2}$ is the identity matrix and $\sigma_i$ are the Pauli matrices:  ${\rho}_\mathbf{p}={1}/{2}[\mathrm{Tr}({\rho}_\mathbf{p}) +{\boldsymbol{{P}}}_\mathbf{p}\cdot\boldsymbol{\sigma}]$. Thus, Eqs.~(\ref{eq:eom1-blocked}) become: 
\begin{subequations}
\begin{alignat}{2}
    i\hbar \frac{\partial \mathrm{Tr}({\rho}_\mathbf{p})}{\partial t} &= 0\, ,\\
    i\hbar \frac{\partial \mathbf{P}_\mathbf{p}}{\partial t} &=\mu 
 (\mathbf{\tilde{D}}_0 -v\mathbf{\tilde{D}}_1 )\times {\boldsymbol{\tilde{{P}}}}_\mathbf{p}\, ,\label{eq:precession}
    \end{alignat}
\end{subequations}
and likewise for antineutrinos. 

\begin{figure}[]
\centering
\includegraphics[width=8cm]{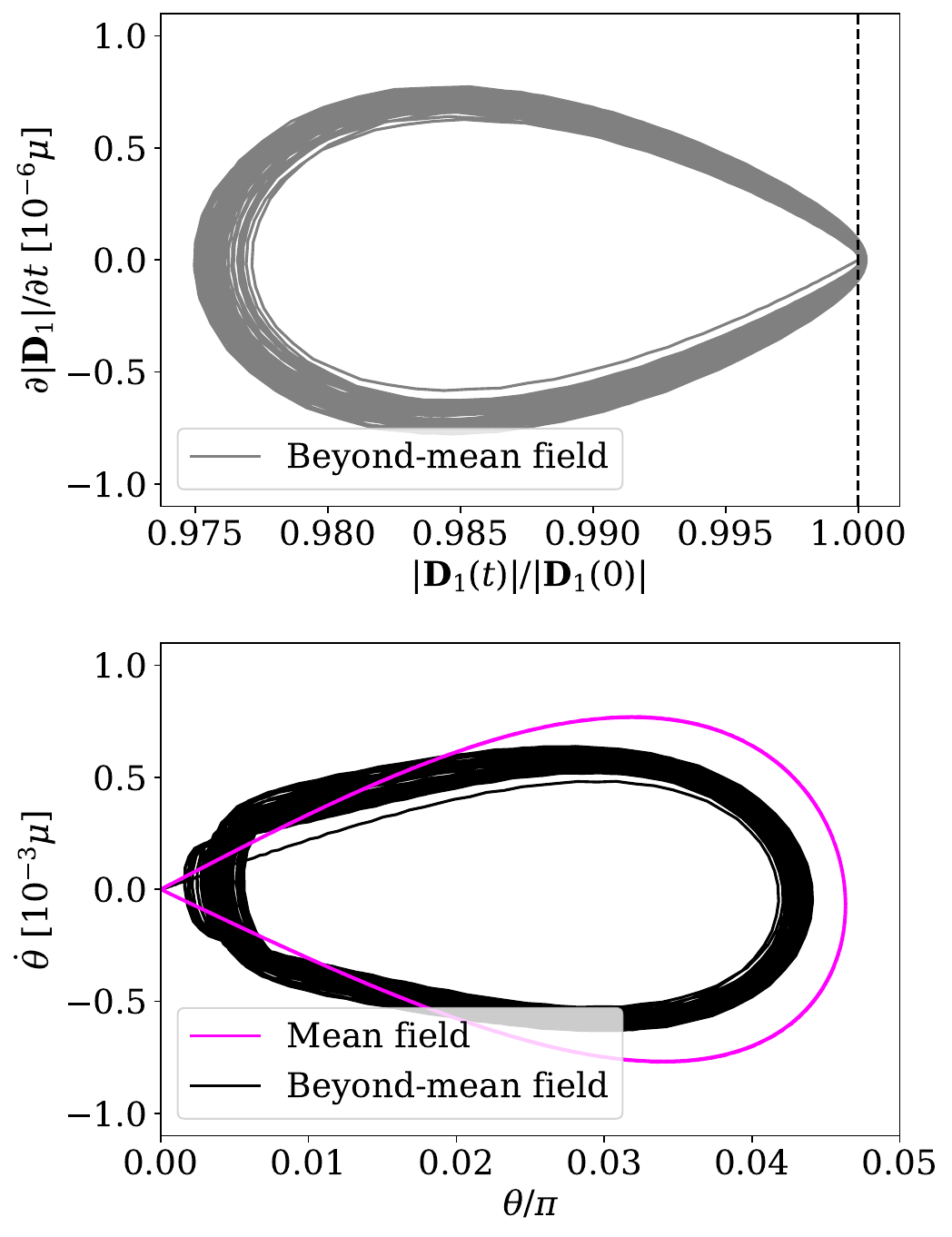}
\caption{Phase-space diagrams of $\mathbf{D}_1$(t) for Case B. \textit{Top panel:} The beyond-mean-field terms change the length of the dipole vector. \textit{Bottom panel:} 
 Flavor conversion in the beyond-mean field case is not periodic, and  $\mathbf{D}_1$ tilts away from the flavor axis over time.
 }
\label{fig:angle}
\end{figure}

Coherent forward scattering, see Eq.~\eqref{eq:cfs}, preserves the number of particles in each momentum mode, so $\mathrm{Tr}(\rho_\mathbf{p})$ is conserved and flavor conversion is driven by $\mathbf{P}_\mathbf{p}$ and $\mathbf{\bar{P}}_\mathbf{p}$. In the presence of neutrino-neutrino interactions, these polarization vectors do not evolve independently of each other.
In the mean-field case, the behavior of the system can be described with a classical analogy: the dipole moment of the ensemble $\mathbf{D}_1$ acts like a gyroscopic pendulum, which precesses instantaneously around a time-dependent vector, $\mathbf{\dot{D}}_1 = \mu (\mathbf{D}_0+\mathbf{D}_2)\times \mathbf{D}_1$. The monopole $\mathbf{D}_0$ plays the role of gravity and a linear combination of the multipoles mimics the angular momentum~\cite{Padilla-Gay:2021haz}. If the ELN distribution is stable, the pendulum is in the ``sleeping top'' configuration. If the ELN distribution is unstable,  bipolar oscillations take place. 
This analogy does not hold when our prescription for neutrino degeneracy is included. As in the mean-field case,  $\mathbf{P}_\mathbf{p}$ and $\bar{\mathbf{P}}_\mathbf{p}$ precess around a certain vector $\mathbf{\tilde{V}}\equiv \mathbf{\tilde{D}}_0 -v\mathbf{\tilde{D}}_1$. However, their $x$ and $y$ components are affected by the beyond-mean-field corrections. When the occupation numbers are low, the latter can be approximated using $(1-B_l)^{1/2}\simeq 1-{1}/{2}B_l$:

\begin{subequations}
\label{eq:length-changing}
\begin{alignat}{2}
    i\hbar \frac{\partial {{P}}^x_{\mathbf{p};bl}}{\partial t}&\simeq  -\frac{\mu}{2}\tilde{V}^y({B}_{e;\mathbf{p}}{\rho}_{ee;\mathbf{p}} -{B}_{x;\mathbf{p}}{\rho}_{xx;\mathbf{p}})\, ,\\
    i\hbar \frac{\partial {P}^y_{\mathbf{p};bl}}{\partial t}&\simeq  +\frac{\mu}{2}\tilde{V}^x({B}_{e;\mathbf{p}}{\rho}_{ee;\mathbf{p}} -{B}_{x;\mathbf{p}}{\rho}_{xx;\mathbf{p}})\, .
    \end{alignat}
\end{subequations}
Here, the upper indices denote the component of the vector in the Cartesian coordinate frame.
The equations for antineutrinos are analogous. 
The terms in Eqs.~\eqref{eq:length-changing} have two distinct effects in the non-linear regime of fast flavor conversion. Firstly, they  change the length of  $|\mathbf{P}_\mathbf{p}|$ and $|\mathbf{\bar{P}}_\mathbf{p}|$, and, if $B_{l;\mathbf{p}}\neq \bar{B}_{l;\mathbf{p}}$, also $|\mathbf{D}_1|$--all of which are conserved in the mean-field case~\cite{Padilla-Gay:2021haz}. Secondly, they cause the evolution of the ensemble to become non-periodic. 
We show these results for our Case B in the top and bottom panels of Fig.~\ref{fig:angle}, respectively. 

The top panel of Fig.~\ref{fig:angle}  describes the time evolution of $|\mathbf{D}_1|$, the length of the dipole vector. 
In principle, the violation of the conservation laws implies that the beyond-mean-field system should have more degrees of freedom than the mean-field one. However, the variation in $|\mathbf{D}_1|$ is small, because Eqs.~\eqref{eq:length-changing} lead to subdominant effects with respect to the mean-field terms. Hence, the symmetries of the mean-field equation of motion are approximately respected. We have attempted to test this using the Gram matrix, which is defined as $G_{mn}=\int_{t_0}^{t_f}\mathrm{d}t\mathbf{P}_{v_m}(t)\cdot\mathbf{P}_{v_n}(t)$ for an arbitrary time interval $(t_0,t_f)$~\cite{Padilla-Gay:2021haz, Raffelt:2011yb}. Here, $v_n$ and $v_m$ denote discrete angular modes. If the Gram matrix has rank $N$, the neutrino ensemble can be mapped into a system of $N$ linearly independent beams. For the examples in Fig.~\ref{fig:R-ee}, we find $N=3$ in both the mean-field and beyond-mean-field cases. Thus, when our beyond-mean-field prescription is taken into account, the system retains a large degree of coherence, even though it does not behave like a gyroscopic pendulum.

The bottom panel of Fig.~\ref{fig:angle} depicts  the evolution of the zenith angle $\theta \equiv \mathrm{tan}^{-1}(\sqrt{(D_{1}^x)^2+(D_{1}^y)^2}/D_{1}^z)$. In both the mean-field and beyond-mean-field cases, the system starts in the upright position ($\theta =0$).
Then, in the mean-field case, flavor conversion creates a cyclic trajectory between $\theta =0$ and a fixed endpoint, which depends on the shape of the angular distribution. 
On the contrary, in the scenario including beyond-mean-field modifications, the evolution is neither periodic nor bipolar. Over time, the interplay between the mean-field and the higher-order terms in Eq.~\eqref{eq:precession} causes the vector $\mathbf{D}_1$  to tilt away from the $z$-axis. 

\begin{figure*}
\centering
\includegraphics[width=\textwidth]{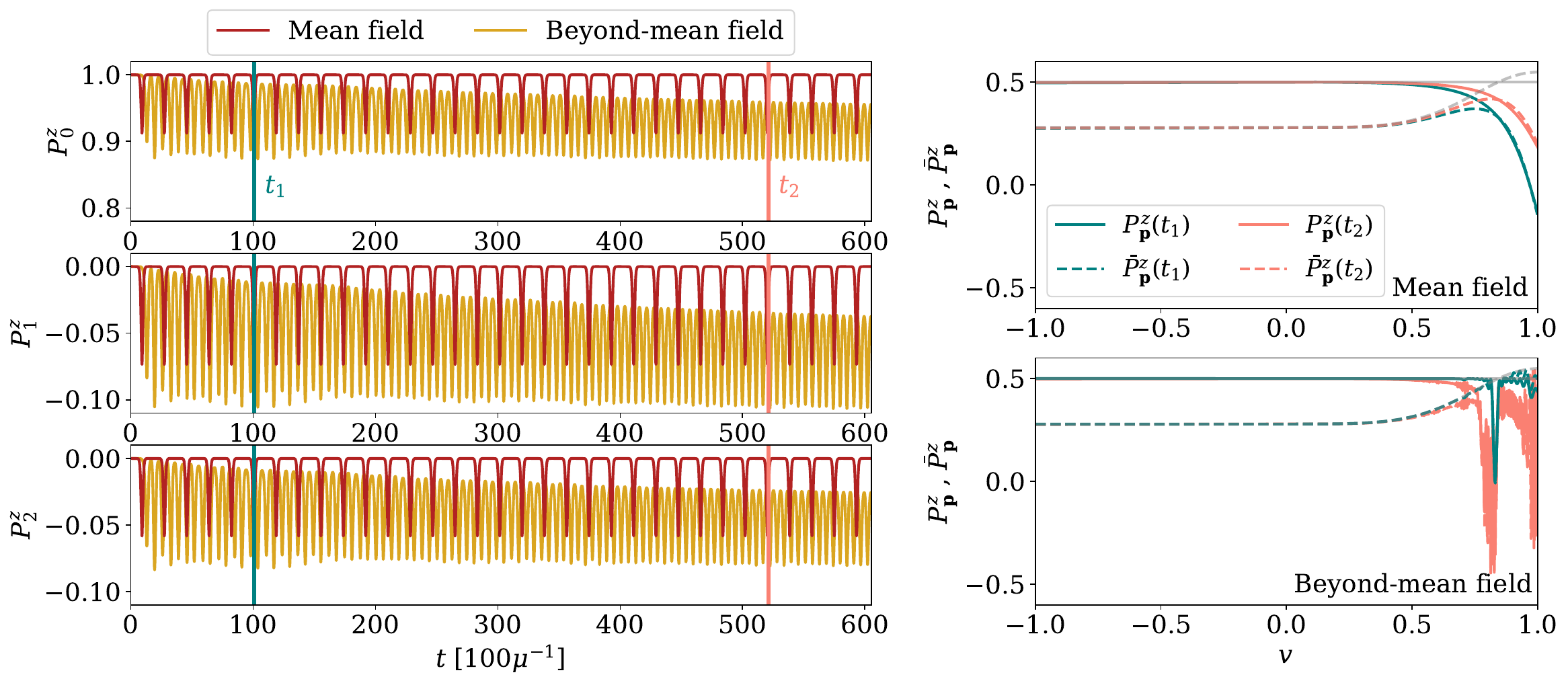}
\caption{\textit{Left panels:}
Time evolution of the multipoles of the neutrino angular distribution up to $n=2$ for Case B. The average value of the multipoles evolves over time, reflecting the cascade of flavor waves towards smaller angular scales. The vertical lines mark the time snapshots shown in the right panels. 
\textit{Right panels:} Angular distribution of the $z$ component of the polarization vectors, $P^z_{\mathbf{p}}$ (solid) and $\bar{P}^z_{\mathbf{p}}$ (dashed), for selected time snapshots in Case B. Gray lines mark the initial conditions. Flavor conversion spreads across angular modes in the case including beyond-mean-field terms, whereas it remains localized around the ELN crossing in the mean-field case. }

\label{fig:poles}
\end{figure*}

{In general, the beyond-mean-field terms in Eqs.~\eqref{eq:length-changing} are different for each $\mathbf{P}_\mathbf{p}$ and  $\mathbf{\bar{P}}_\mathbf{p}$, since each polarization vector follows its own equation of motion. Adopting a decomposition in multipoles, e.g. $\mathbf{P}_\mathbf{p}=\sum_{n=0}^\infty (n+\frac{1}{2})\mathbf{P}_n$, one can see a cascade of flavor instabilities towards ever smaller angular scales (cf.~also Ref.~\cite{Johns:2020qsk}), which can be described as the growth of higher-order multipoles $P_n^z$ and $\bar{P}_n^z$ (see Fig.~\ref{fig:poles}, left panels). As a visual aid, the right panels in Fig.~\ref{fig:poles}  highlight the effect of the instability on the neutrino and antineutrino angular distributions for selected time snapshots in Case B. The top right panel shows the fast flavor pendulum: angular modes in the proximity of the  ELN crossing evolve, undergoing synchronous oscillations. The bottom right panel shows that, when beyond-mean-field corrections are included, flavor conversion begins around the ELN crossing, but spreads across angular modes.} 

{We note that the equivalence between the $P_\mathbf{p}^z$ and $\bar{P}_\mathbf{p}^z$ equations of motion can be broken within the mean-field approximation.  Indeed, if Eqs.~\eqref{eq:eom1} are extended to include vacuum oscillations, the $P_{n>0}^z$ and $\bar{P}_{n>0}$ multipoles  also grow: see e.g.~Ref.~\cite{Shalgar:2020xns}. Incoherent scattering can also lead to a similar outcome, if the collision terms violate the symmetries of Eqs.~\eqref{eq:eom1}: see e.g.~Ref.~\cite{Shalgar:2020wcx}. For example, fast flavor instabilities cascade to small angular scales when the scattering terms are different for neutrinos and antineutrinos or, more generally, when they are energy-dependent. Even if the scattering rates are equal, collisions break the periodicity of flavor evolution, and cause a decrease in $|\mathbf{D}_1|$~\cite{Raffelt:2011yb, Padilla-Gay:2022wck}. Hence, several different effects could lead to the same effective flavor outcome.} 

\begin{figure*}
\centering
\includegraphics[width=\textwidth]{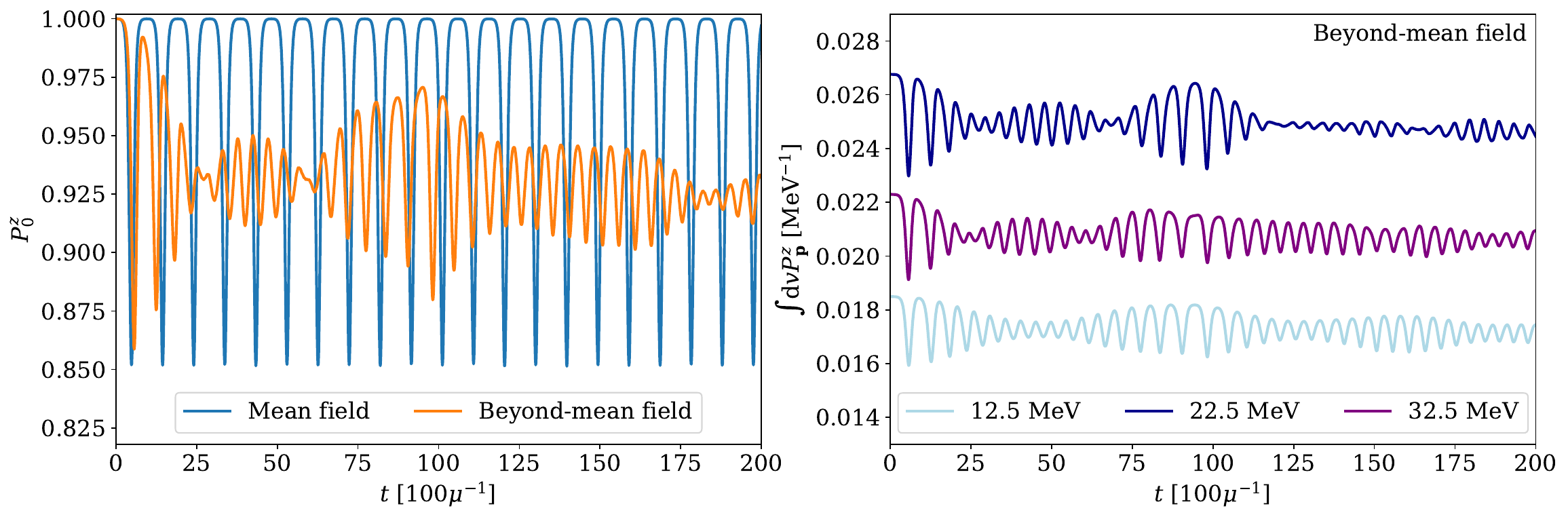}
\caption{\textit{Left panel:} Monopole of the neutrino distribution as a function of time, for Case B. In the beyond-mean-field scenario, flavor conversion does not exhibit a bipolar behavior, due to the energy dependence of the additional terms in the equation of motion. \textit{Right panel:} Time evolution of selected angle-integrated neutrino energy modes. Each polarization vector decoheres from the others, changing the shape of the energy spectrum.} 
\label{fig:spectrum}
\end{figure*}

\section{Energy-dependent features of beyond-mean-field effects}
\label{sec:multi-energy}

In order to investigate whether the beyond-mean-field corrections induce energy-dependent effects, we solve Eqs.~\eqref{eq:eom1} and \eqref{eq:eom1-blocked} assuming Fermi-Dirac distributions for neutrinos and antineutrinos, with the initial distribution in Eq.~\eqref{eq:ang}. We consider $T=10$~MeV, $\mu_{\nu_e}=3$~MeV, and the same $b$ and $d$ as in Case B. 
For the sake of simplicity, we neglect vacuum mixing to highlight the effect of the beyond-mean-field corrections and generate flavor coherence by including a small off-diagonal perturbation in the initial conditions.

The left panel of Fig.~\ref{fig:spectrum} displays the energy- and angle-integrated (anti)neutrino distribution, $P_0^z$ ($\bar{P}_0^z$), as a function of time. The mean-field system still behaves as a gyroscopic pendulum, because the self-interaction term is energy-independent. {In the beyond-mean-field case, each polarization vector follows a different equation of motion, due to the additional terms in Eq.~\eqref{eq:length-changing}: hence, flavor evolution is no longer bipolar. Rather, each energy mode decoheres from the others, as shown in the right panel of Fig.~\ref{fig:spectrum}. This means that beyond-mean-field terms can change the shape of the neutrino spectra, and the behavior of the multi-energy ensemble is intrinsically different from that of a single-energy approximation.}
{However, the evolution of neutrinos and antineutrinos is still identical, which is a consequence of self-interactions conserving the ELN. Indeed, integrating over $v$ and $E$ in Eq.~\eqref{eq:precession} and the analogous expression for antineutrinos leads to  
    $i\hbar {\partial \mathbf{D}_0}/{\partial t} =\mu (\mathbf{\tilde{D}}_0\times \mathbf{\tilde{D}}_0-\mathbf{\tilde{D}}_1 \times \mathbf{\tilde{D}}_1) =0$.
}

\section{Outlook}
\label{sec:end}
The physics of flavor conversion due to neutrino-neutrino interactions shows a very rich phenomenology that we are still far from fully grasping. Most of the existing work on this topic is carried out employing the mean-field approximation. In this paper, we have attempted to investigate whether beyond-mean-field effects on neutrino-neutrino coherent forward scattering could introduce flavor instabilities that are not otherwise foreseen. To this end, we have focused on the regime where neutrinos are degenerate and  
introduced a heuristic prescription for Pauli-blocking terms in the neutrino self-interaction Hamiltonian. 
Such corrections define an effective ELN distribution, which modifies the (anti)neutrino dispersion relation. Hence, ensembles that are on the verge of stability in the mean-field case can become unstable when the beyond-mean-field terms are included and vice versa. 

Relying on the gyroscopic pendulum analogy introduced for fast flavor conversion~\cite{Padilla-Gay:2021haz, Johns:2019izj}, we have investigated how the non-linear regime of the flavor evolution is modified by our beyond-mean-field prescription. Within the mean-field approximation, each (anti)neutrino polarization vector precesses around the potential generated by all other neutrinos in the ensemble.  In the beyond-mean-field scenario, additionally, the length of the polarization vectors changes due to dynamical decoherence. The terms that cause this effect introduce an energy dependence in the collective potential and break the symmetry between the neutrino and the antineutrino equations of motion. As a result, flavor instabilities cascade down to small angular scales, and the evolution of the system is not periodic anymore.

Qualitatively, the outcome of flavor conversion  including the beyond-mean-field effects in our toy model is similar to what is observed when incoherent scattering and vacuum mixing are included in the mean-field equations of motion, {although the physics linked to each of these phenomena and their interplay with the self-interaction term in the Hamiltonian are different}. We have neglected these additional terms in the equations of motion throughout this work, to highlight the impact of beyond-mean-field corrections on fast instabilities. However, collisions, vacuum mixing, and potential beyond-mean-field corrections (due to neutrino degeneracy) should  coexist in the  core of an astrophysical system.

\acknowledgments
We are grateful to Georg Raffelt for useful discussions. This project has received support from the Villum Foundation (Project No.~13164), the Danmarks Frie Forskningsfond (Project 
No.~8049-00038B), the European Union (ERC, ANET, Project No.~101087058), and the Deutsche Forschungsgemeinschaft through Sonderforschungbereich SFB 1258 ``Neutrinos and Dark Matter in Astro- and Particle Physics'' (NDM). 
Views and opinions expressed are those of the authors only and do not necessarily reflect those of the European Union or the European Research Council. Neither the European Union nor the granting authority can be held responsible for them. The Tycho supercomputer hosted at the SCIENCE HPC Center at the University of Copenhagen was used for supporting the numerical simulations presented in this work.

\appendix

\section{Correlation functions and scattering amplitudes}
\label{sec:appendix}

\begin{figure}[H]
\centering
\includegraphics[width=9.5cm]{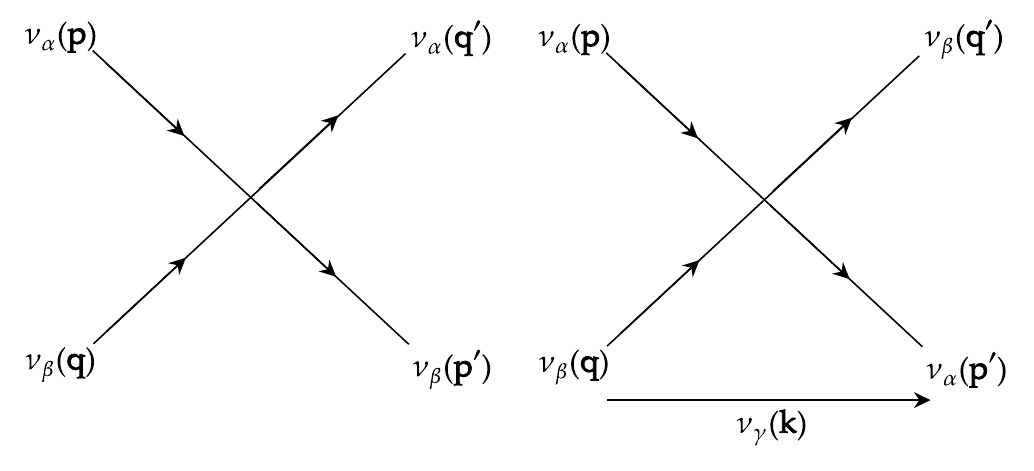}
\caption{\textit{Left:} Two-particle neutrino-neutrino interaction in the low-energy approximation. \textit{Right:} Same as left panel,  but in the presence of a spectator neutrino $\nu_\gamma (\mathbf{k})$. These diagrams can describe the same processes sketched in Fig.~\ref{fig:sketch-bmf}, setting $a = e$, $\gamma =b=x$.}
\label{fig:diagram}
\end{figure}

In neutrino-dense astrophysical sources,  neutrino-neutrino scattering can be modeled as a Fermi interaction, as shown in the left panel of Fig.~\ref{fig:diagram}. The external lines represent the neutrino field $\nu_\alpha (\mathbf{p})=a_\mathbf{p}u_\mathbf{p}e^{-i\mathbf{p}\cdot\mathbf{r}}$, where $u_\mathbf{p}$ is the left-handed particle spinor. The amplitude of the process $\nu_\alpha (\mathbf{p})+\nu_\beta (\mathbf{q})\to \nu_\alpha (\mathbf{p'})+\nu_\beta (\mathbf{q'})$ can be written as:
\begin{equation}
\label{eq:scatt-amp}
\begin{array}{rl}
H_{\alpha\beta }(\mathbf{p'},\mathbf{q'};\mathbf{p},\mathbf{q})  & = \mu  [\bar{u}_\mathbf{p'}\gamma^\mu P_Lu_\mathbf{p}][\bar{u}_\mathbf{q'}\gamma_\mu P_Lu_\mathbf{q}]\\ & \times a^\dagger_{\alpha ,\mathbf{p'}}a_{\alpha ,\mathbf{p}}a^\dagger_{\beta ,\mathbf{q'}}a_{\beta ,\mathbf{q}} \,,
\end{array}
\end{equation}
where $P_L\equiv (1-\gamma_5)/2$.
In dense environments, the two-particle scattering processes occur in the presence of  other neutrinos that act as ``spectators'' (cf.~the right panel of Fig.~\ref{fig:diagram}). The latter do not participate in the interaction; since particle spinors are normalized such that $\bar{u}_\mathbf{p}u_\mathbf{p}=1$, their contribution to the diagram is: 
\begin{equation}
    H^{(3)}_{\alpha\beta\gamma}(\mathbf{p'},\mathbf{q'},\mathbf{k};\mathbf{p},\mathbf{q},\mathbf{k})=H_{\alpha\beta}(\mathbf{p'},\mathbf{q'};\mathbf{p},\mathbf{q})a^\dagger_{\gamma,\mathbf{k}}a_{\gamma ,\mathbf{k}}\,.
\end{equation}

In this paper, we aim to investigate whether,  in degenerate environments, this additional pair of creation/annihilation operators can have important effects on the development of flavor instabilies. At the mean-field level, the evolution of the single-particle density matrix in Eq. \eqref{eq:rho} is given by:
  \begin{equation}
  \label{eq:Heisenberg}
      i\dot{\rho}_{\mathbf{p},\gamma\kappa} =\langle [a^\dagger_{\kappa,\mathbf{p}}a_{\gamma,\mathbf{p}},H]\rangle\,,
  \end{equation}
where brackets denote an ensemble average. To describe the effect of neutrino-neutrino scattering, the Hamiltonian $H$ must be of the form shown in Eq. \eqref{eq:scatt-amp}, integrated over $\mathbf{p'}$, $\mathbf{q'}$ and $\mathbf{q}$. Wick's theorem can be used to decompose the right-hand side of the equation into products of two-operator correlation functions. For example, for $\kappa =\gamma =\alpha =e$ and $\beta = x$, this yields terms such as: 
  \begin{equation}
  \begin{array}{rl}
  \label{eq:wick}
      \langle a^\dagger_{e,\mathbf{p'}}a_{e,\mathbf{p}}a^\dagger_{x,\mathbf{q'}}a_{x,\mathbf{q}}a^\dagger_{e,\mathbf{p}}a_{e,\mathbf{p}}\rangle  & = \langle a^\dagger_{e,\mathbf{p'}}a_{e,\mathbf{p}}a^\dagger_{x,\mathbf{q'}}a_{x,\mathbf{q}}\rangle \,\\ &= \langle a^\dagger_{e,\mathbf{p'}}a_{e,\mathbf{p}}\rangle\langle a^\dagger_{x,\mathbf{q'}}a_{x,\mathbf{q}}\rangle \\ &+\langle a^\dagger_{e,\mathbf{p'}}a_{x,\mathbf{p}}\rangle\langle a_{e,\mathbf{p}}a^\dagger_{x,\mathbf{q'}}\rangle \,, \end{array} 
  \end{equation}
  where the first equality is given by the anti-commutation relation $a^\dagger_{e,\mathbf{p}}a_{e,\mathbf{p}} =1-a_{e,\mathbf{p}}a^\dagger_{e,\mathbf{p}}$. Since the mean-field approximation forbids correlations between different momentum modes, i.e. $\langle a^\dagger_{\alpha,\mathbf{p'}}a_{\beta,\mathbf{p}}\rangle =(2\pi)^3\delta (\mathbf{p'}-\mathbf{p})\rho_{\mathbf{p},\beta\alpha} $, all contributions vanish unless $\mathbf{p'}=\mathbf{q}$ and $\mathbf{q'}=\mathbf{p}$ (``forward scattering'') or $\mathbf{p'}=\mathbf{p}$ and $\mathbf{q'}=\mathbf{q}$ (no scattering). Thus, Eq. \eqref{eq:Heisenberg} provides a closed system of differential equations,  describing neutrinos with fixed momenta propagating in an external potential. However, this approximation relies on the assumption that the neutrino gas is   homogeneous, which is not an accurate picture of neutrino ensembles in dense astrophysical systems. Momentum correlations of the form $\langle a^\dagger_{\beta,\mathbf{p}}a_{\alpha,\mathbf{q}}\rangle$ must exist in these environments, e.g.~due to spatial inhomogeneities. It is not clear whether two-particle scattering amplitudes are enough to model quantum kinetics. Indeed, applying Wick's theorem to the ensemble average $\langle H^{(3)}_{exe}\rangle$ yields:
  \begin{equation}
  \label{eq:3-amplitude}
      \begin{array}{rl}
\langle H^{(3)}_{exe}\rangle &= \langle H_{ex}\rangle \langle a^\dagger_{e,\mathbf{k}}a_{e,\mathbf{k}}\rangle   -\langle a^\dagger_{e,\mathbf{q'}}a_{e,\mathbf{k}}\rangle(\langle a^\dagger_{x,\mathbf{p'}}a_{x,\mathbf{q}}\rangle\\ &\times \langle a^\dagger_{e,\mathbf{k}}a_{e,\mathbf{q}}\rangle+\langle a^\dagger_{x,\mathbf{p'}}a_{e,\mathbf{p}}\rangle\langle a^\dagger_{e,\mathbf{k}}a_{x,\mathbf{p}} \rangle )+\langle a^\dagger_{x,\mathbf{p'}}a_{e,\mathbf{k}}\rangle  \\ & \times (\langle a_{x,\mathbf{p}}a^\dagger_{e,\mathbf{q'}}\rangle \langle a^\dagger_{e,\mathbf{k}} a_{e,\mathbf{q}}\rangle +\langle a^\dagger_{e,\mathbf{q'}}a_{e,\mathbf{p}}\rangle \langle a^\dagger_{e,\mathbf{k}}a_{x,\mathbf{p}}\rangle )\,.
      \end{array}
  \end{equation}
  
  The additional terms provided by the neutrino acting as a spectator may affect the two-particle result. {Let us consider $\alpha = e$ and $\beta =\gamma=x$, i.e., $\nu_e(\mathbf{p})+\nu_x(\mathbf{q})+\nu_x(\mathbf{k})\to \nu_e(\mathbf{p'})+\nu_x(\mathbf{q'})+\nu_x(\mathbf{k})$. Processes of this form are Pauli-blocked if $\mathbf{q'}=\mathbf{k}$. As argued in Sec.~\ref{sec:bmf}, these interactions can be coherently enhanced if momentum transfer is small. Hence, neutrino refraction should be sensitive to the presence of degenerate spectator particles. The scenario sketched in Fig.~\ref{fig:sketch-bmf} corresponds to $\alpha =e$,  $\beta =\gamma =x$ and $\mathbf{p'}=\mathbf{k}$. This process is allowed by the exclusion principle, but it cannot increase the flavor coherence of the incoming $\nu_e(\mathbf{p})$. In a physical setting, the outgoing particle would be in flavor superposition, and thus indistinguishable from the spectator, so the final state of the interaction must be anti-symmetric in these two particles. If their final momenta are also identical ($\mathbf{p'}\sim \mathbf{k}$), the only possibility is that they become a singlet, so that they always occupy a different flavor eigenstate.} 
  
  {To account for this effect in a kinetic approach, it would be necessary to calculate the correlations $\langle a^\dagger_{e,\mathbf{p}}a_{x,\mathbf{q}}\rangle$ for all $\mathbf{p}$ and $\mathbf{q}$. This would require a self-consistent way to solve the higher-order equations in the BBGKY hierarchy, which is beyond the scope of this work. Instead, } we assess whether new features  could be introduced  in flavor conversion by  momentum correlations {with spectator particles}. For this purpose,  we  assume that they are responsible for  a correction proportional to the neutrino occupation number in the refractive term of the quantum kinetic equation. 

\bibliography{blocking_conversion.bib}

\end{document}